# Go Together: Bridging the Gap between Learners and Teachers

**Asim Irfan**
Computer Systems Engineering
Mehran University of Engineering and Technology, Pakistan

**Atif Nawaz**
Computer Systems Engineering
Mehran University of Engineering and Technology, Pakistan

**Muhammad Turab**
Computer Systems Engineering
Mehran University of Engineering and Technology, Pakistan

**Muhmmad Azeem**
Computer Systems Engineering
Mehran University of Engineering and Technology, Pakistan

**Mashal Adnan**
Design Internee UI/UX
UnifiedCrest R&D Group

**Ahsan Mehmood**
Founder Psychologist
(Helping Mind)

**Sarfaraz Ahmed**
Co-founder Psychologist
(Helping Mind)

**Dr. Adnan Ashraf**
Faculty, CS, MUET, Jamshoro

*Abstract*—After the pandemic, humanity has been facing different types of challenges. Social relationships, societal values, and academic and professional behavior have been hit the most. People are shifting their routines to social media, and gadgets and getting addicted to their isolation [5]. This sudden change in their lives has caused an unusual social breakdown and endangered their mental health. In mid of 2021, Pakistan's 1st Human Library was established under HelpingMind® to overcome these effects [9]. Despite online sessions and webinars, HelpingMind® needs technology to reach the masses. In this work, we customized the UI/UX of a Go-Together Mobile Application (GTMA) to meet the requirements of the client organization. A very interesting concept of the book (expert listener/ psychologist) and the reader is introduced in GTMA. It offers separate dashboards, separate reviews or rating systems, booking, and venue information to engage the human-reader with his/ her favorite human book. The loyalty program enables the members to avail discounts through a mobile application and its membership is global where both the human-reader and human-books can register under the platform. The minimum viable product has been approved by our client organization.

*Keywords—Go-Together Mobile Application (GTMA), UI/UX, Pandemic, Pakistan, HelpingMind (HM),*

## I. Introduction

The post-pandemic era coined a new meaning to human life that carries changed routines with digital devices and social distancing [11],[25]. People, institutes, organizations, and companies shifted their work to online platforms and that phase of digitalization and social distancing has negatively affected people's mentality [5],[26],[27],[28]. To overcome this issue, the need for psychologists, practitioners, doctors, and experienced brains has increased more than ever before. An idea is proposed to make people find the people of their interest. The project is about the design and development of a mobile application for a startup Human Library, a library having live books in every known field. People can contact experts of their desire from different fields and are known as Live Books meeting with the Reader who requests the expert under an endorsed mutual consent. This encompasses a novel way of giving and taking appointments, and calendar sharing. The Human Library Application project may or will bring all Live Books together for their Readers. This process ensures that the benefit from one person to another is direct and everyone shares their best under the shared economy.

The application is also known as Go-Together Mobile Application (GTMA) throughout this document. Now, coming to the working mechanism of GTMA. It starts with the login page. When the application is opened a visitor shall land on a screen having two options. The first one is to explore HelpingMind (HM) as a guest and the second one is to sign-up as a New User by clicking on the "sign-up" button. A user can sign-up as a book or as a reader. Moreover, the information required for sign-up includes Email, First Name, Last Name, City, and Country. If a user is signing-up as a book, the user will be able to browse courses, events, etc. And can generate new events requests but if the user is signing-up as a reader then it will let the user just to browse courses and events only. The user can also enroll/ pay as per the availability of seats and location. Browse the public information including courses and events but, whenever the visitor clicks to generate/enroll in an event, the sign-up information form or page shall appear. A visitor starts exploring HelpingMind (HM) as a guest and then wants to participate in any other activity then the user has to sign-up as well.

At the time of exploring HM as a guest user will not be able to perform any activity but can only see the event details, view the book's profile and can read the book's reviews. To sign-up for performing any activity, user is required to fill their Name, Email, Contact Number, Password, and Re-enter Password. There is also an option, forgot password, in case the user forgets the password and tries to access the account then by tapping on" Forgot Password" the user will be redirected to a new screen in which the user has to enter the email which is linked to the user's ID. An auto-generated OTP will be mailed to the user's linked Email address and the user then can create a new password after entering that OPT in the given field. Now, the login process is completed, and the user has been provided with a home screen where the user can search for a new activity or a session by using the "search option" given at the top of the home screen. Searching will filter those events that the user has searched for and also show related events. The BOOK can be searched by name or by profession. There will be options i.e. All, Events, and Private session through which the user can specify the category. If a user taps on any of the options search results will be filtered out accordingly.

The sidebar is the most important global element and has the most features. Every feature leads to a different screen. As the user taps on the sidebar button, the options My Profile, Messages, Calendar, Payment Method, Vaccination Card, Help, Settings, FAQ, Events near you, and Logout will be displayed. Each option contains further options to complete an activity. The option My profile contains all the information about the user, the user can write and update his personal information at any time and the user's information will be stored on the admin dashboard.

The user can send the message to the event management or any book for a private session and can see the message's history. The messages can be sent to only those books whom the user follows. The visitor will be able to see all the events which are going to happen in the coming time. The dates of events will automatically be highlighted. The visitor is exploring the event or session and if the visitor wants to attend that session, then he/she must book a seat. The user can book seats by online payment through Easypaisa or Jazzcash as these are easy, simple and mostly used payment methods in Pakistan.

The requirement includes a vaccination card for any activity to be performed as the recent attack of coronavirus has caused death to millions of people around the globe.

So, it is very much important for everyone to ensure that they are safe to meet through the vaccination card. Every user of the application must be vaccinated and must upload scanned photos of the vaccination card's both sides. Any user who needs any help regarding the application can contact an admin through email, or contact number or can directly approach the given address. There is another option of settings which further includes five options i.e. Account, Notifications, Privacy, Help & support, and about us. The user can navigate to the settings option, then account option to change the personal information of his/her account and can change passwords as well, using the privacy option. For the users interested in attending any session, has to book a seat in the session by moving to the seat booking page where the they will be able to see Session Name, Session date, Session Address, and Session Location on google Maps.

Moreover, to become a BOOK, user would need to tap on the button (Become a book) available on the home screen and the user will be redirected to a screen where the following details: Name, Phone, CNIC, Field, Vaccination Card (image) and Resume/CV (file) must be filled. After filling out this form, a request will be sent to the admin and the user will have to wait until the admin accepts the request and verifies whether the user completes the requirement of becoming a book or not. The user will receive an email from the admin about their request to become a book has been accepted or rejected.

There is another essential element: notifications, which plays a vital role for users, as users will be getting all information about new events or new activities through the notifications. For example, the management team posts about their event on the calendar every user will receive a notification about the event. Readers will also receive notifications from BOOKS, which they follow, whenever they post about free time on the calendar.

## II. LITERATURE REVIEW

### A. Zoik IT

Zoik IT is a mobile application designed and developed by Zoik Technologies Pvt. Ltd. The purpose of the application is not specified, but it can be assumed that it is a general-purpose IT app aimed at providing solutions to various information technology-related challenges. The app is likely to offer a range of features and functions that help users with various IT-related tasks, such as data management, networking, software development, and more. The app may be targeted toward individuals, small businesses, and enterprise organizations, and its features are designed to meet the needs of each user group. The app is likely to be available for download on the App Store for iOS devices and the Google Play Store for Android devices. The design and development of the app are likely to have considered various factors, such as user experience, performance, and security, to ensure that it provides a seamless and secure user experience [1],[29],[30].

### B. Dil Ka Rishta (DKR)

DilKaRishta is a matrimonial app designed and developed to help individuals find suitable partners for marriage. The app is likely to offer a range of features and functions that help users find potential partners based on their preferences, such as age, religion, location, education, and more. The app may also provide users with tools for communication and information sharing, such as messaging and profile sharing, to help users get to know each other better. The app is likely to be targeted toward individuals seeking to find a life partner, and its features are designed to meet the needs of this user group. The design and development of the app are likely to have considered various factors, such as user experience, privacy, and security, to ensure that it provides a safe and secure platform for users to find potential partners [2],[31][32].

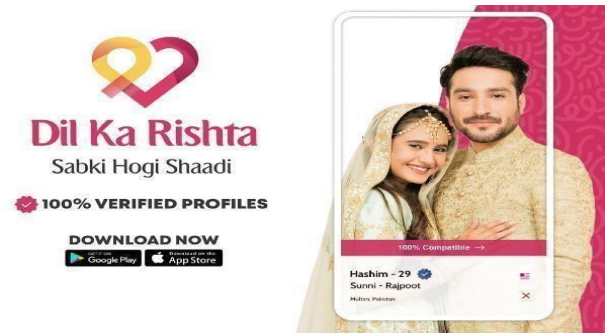

Fig. 1. DilKaRishta Mobile Application.

### C. Bookme.pk

Bookme.pk is an online platform designed and developed to help users book various services and activities in Pakistan. The platform likely offers a wide range of services, such as movie tickets, event tickets, travel packages, and more. The platform is designed to be user-friendly and provide an easy and convenient way for users to book services and activities online. Bookme.pk is likely to be available through its website or as a mobile app on the App Store. The platform's features and functions are designed to provide users with a seamless booking experience and offer a variety of payment options for added convenience. The design and development of the platform are likely to have considered various factors, such as user experience, security, and performance, to ensure that users have a smooth and reliable booking experience [3],[33],[34].

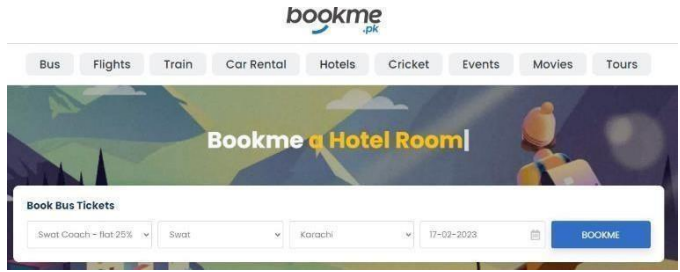

*Fig. 2.Bookme.PK website example image*

### A. Pakistan Railways

There is an application Named Pak-railways that provides seat bookings. It makes travel easy and efficient. People traveling anywhere within Pakistan can book seats to their desired destination and can track the train. It provides the feasibility to leave home knowing the current location of the train. It saves time and energy for humans traveling through the railways. Pakistan railways official is the finest plate form of Pakistan railways offering Pakistan's public the perfect and the most reliable manner to e-book their tickets to travel on the train [3],[35],[36]. Pakistan Railways provides a prime means of transportation to remote corners of the country and brings everyone closer to business, tourism, pilgrimage, and education. It has been a considerable integrating force and forms the lifeline of the country by fulfilling its needs for a mass movement of people and transportation of goods [3],[37].

### B. Eventbrite

A platform for managing events online and selling tickets is called Eventbrite [4]. It gives attendees a simple and practical method to find and register for events while enabling event organizers to plan, advertise, and sell tickets to their events. The platform offers a variety of features and functions, such as event creation and promotion, ticket sales, and check- in management, to help event organizers streamline the event planning process. The platform's user-friendly design, flexible pricing options, and integration with social media make it a popular choice among event organizers of all sizes, from small local events to large-scale international events. The platform's design and development have considered various factors, such as user experience, security, and performance, to ensure that both event organizers and attendees have a seamless and efficient experience. Eventbrite also provides analytics and insights to help event organizers measure the success of their events and make informed decisions for future events [4].

## III. METHODOLOGY

### A. SCRUM

Agile Scrum is a project management framework used to develop and deliver high-quality software. It is based on Agile principles and is designed to help teams work together to produce working software incrementally and iteratively. Scrum events (Sprint, Sprint Planning, Daily Scrum, Sprint Review, and Sprint Retrospective), Scrum roles (Product Owner, Scrum Master, and Development Team), and Scrum artifacts are the core components of the Scrum process (Product Backlog, Sprint Backlog, and Increment). This strategy emphasizes delivering the greatest business value in the shortest amount of time while encouraging cooperation, adaption, and continual development.

Scrum can be applied to the development of mobile applications, as it provides a flexible and adaptable framework for delivering software products. The Scrum process can be used to manage the complexity and uncertainty of mobile app development, by breaking it down into smaller, manageable pieces.

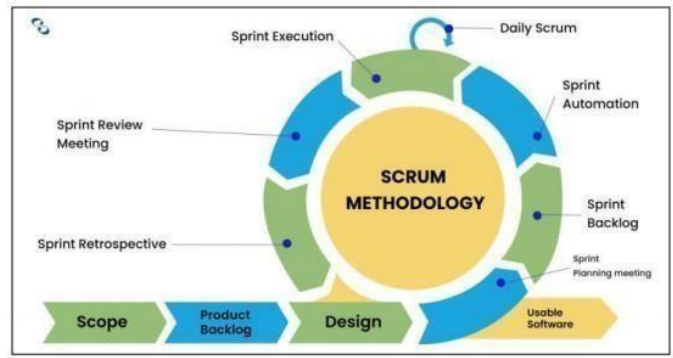

*Fig. 3.Illustration of Scrum methodology*

A prioritized list of the features and requirements for the mobile app is defined by the Product Owner in a Scrum project. In each sprint, which lasts between two and four weeks, the development team then strives to provide a potentially release-ready increment of the program.

The Development Team meets for a 15-minute stand-up meeting each day to discuss progress, identify any roadblocks, and make goals for the following 24 hours. The Sprint Review gives stakeholders a chance to evaluate the most recent app increment and offer comments, whereas the Sprint Retrospective gives the team a chance to look back on the sprint and pinpoint areas for improvement.

By following the Scrum process, teams can rapidly deliver high-quality mobile apps that meet the needs of their users.

### B. Use Case Diagram

Use case charts are a pictorial portrayal of any potential collaborations of the client with the application or framework. A utilization case graph shows numerous or different use cases and numerous or various sorts of clients the application or framework has and frequently will utilize the application or framework. The utilization cases are addressed by either ovals or circles.

In the following Figure, where this use case of a go- together mobile application is more defined regarding the operations performed within the application, and how the components get to work altogether. This use case depicts the user authentication, user registration, and login process at first,to get into the home page. Where all features are ready to availby users. After the registration application user's information is sent to the admin panel for the admin, to know about the user and whether it is valid or not. After the permission of the admin user will be able to login into the account by using the username and password.

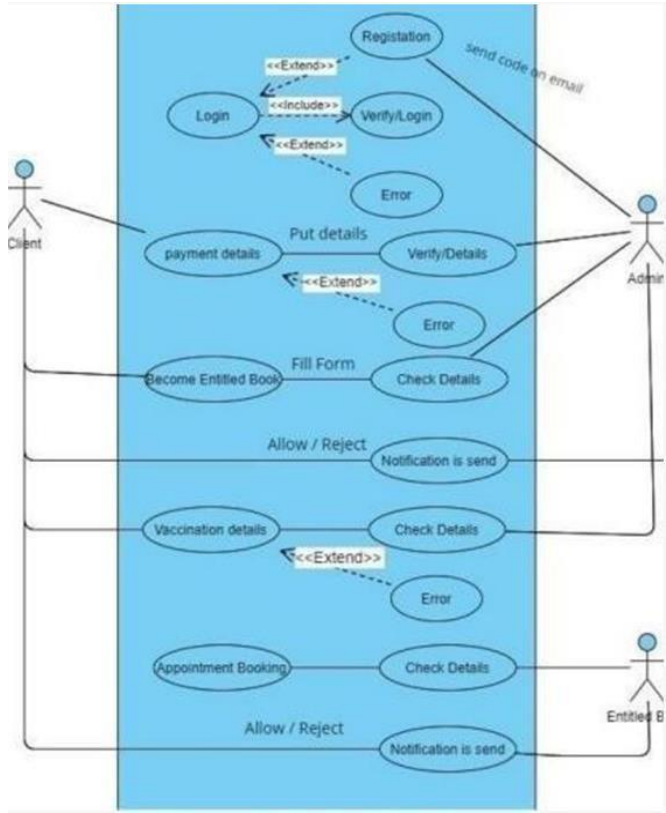

Fig. 4. Proposed Use Case Diagram

*C. Toolkits*

We are going to use the toolkits that are mentioned in the below table for the wireframes, design, and mobile application development.

| Domain | Tools |
|---|---|
| Wireframes and Use-Cases | Balsamiq and Visual Paradigm |
| Designing and Prototype | Adobe XD |
| Coding and Development | Flutter and flutter pro |

We have used Balsamiq to create the Wireframes of the application, as well as we have also used the Visual Paradigm online tool to design the usecases of our application. Talking about the design of the application, we have used Adobe software called Adobe XD (Experience Design) to design our application's layout and userinterface. We have made a very user-friendly userinterface, so that people with lesser knowledge of technology, or people who don't know how to use smartphones, will also find our application very easy to be used and navigate through the application very easily without having to face any difficulties.

We have used Flutter because it is a hybrid mobile application development platform, meaning we can create Android and iOS applications using a single codebase. It is based on a programming language called Dart. We have used its different widgets and packages to develop our application's front-end design. We have done its backend development on Firebase.

Our application's backend development was completed using Firebase as a Backend-as-a-Service (Baas). It offers a variety of tools and services to developers so they may create a wide range of applications, as well as generate high-quality apps, expand their user base, and make money. On the infrastructure of Google, Firebase is constructed. Additionally, it is characterized as a NoSQL database application that stores documents that resemble JSON.

*D. Prototype*

In the following figure, where all neighbors and home screens are displayed. On the home screen, there are event, meeting, session, and class booking options are displayed. This is the procedure of booking the meeting with a specific book.

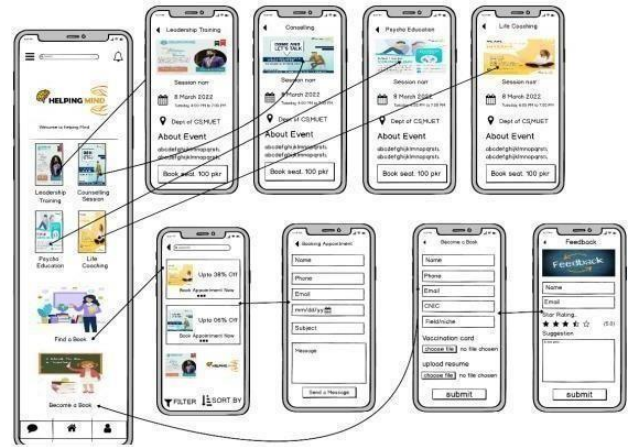

Fig. 5. Home screen and its neighbor screens

IV. RESULTS

*A. Sign Up Screen*

Information required for sign-up includes some basic information. After giving all these details user is ready to use the HM and can perform any activity.

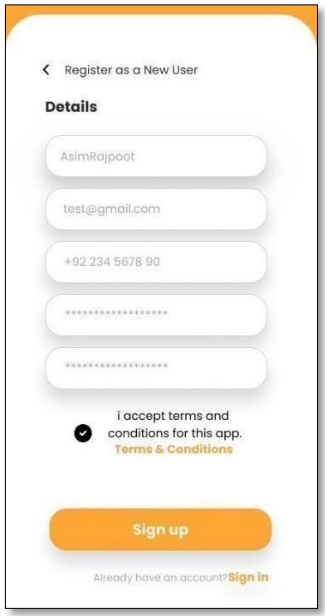

Fig. 6. Sign up Screen of the proposed app

*B. Sign in Screen*

On the sign-in screen, the user can login through his email and the password he creates himself also that user can sign in with google Gmailand with apple id.

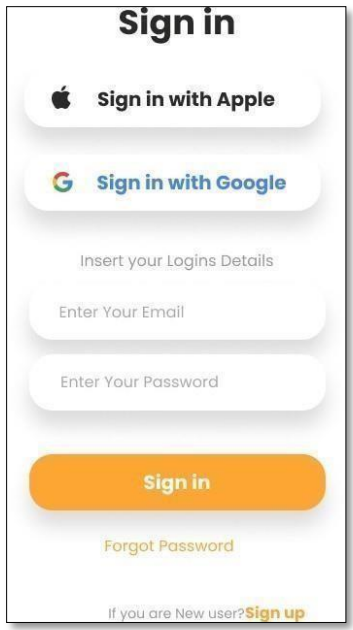

Fig. 7. Sign in Screen of the proposed app

*C. Home Screen*

This is the home screen of the application where users can see all the events organized by Helping Mind and can search for any event. It also includes a sidebar button through which a user can access a lot of other activities.

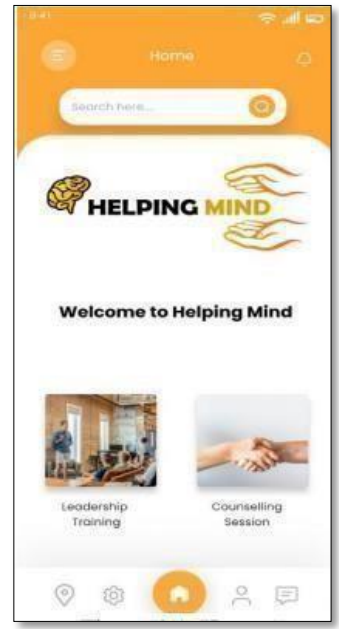

Fig. 8. Home Screen

*D. Payment Method*

A visitor exploring the event or session and wants to attend that session then the user has to book a seat for himself/herself. Users can book a seat online by paying through Easypaisa and JazzCash, as these are the easiest methods to pay.

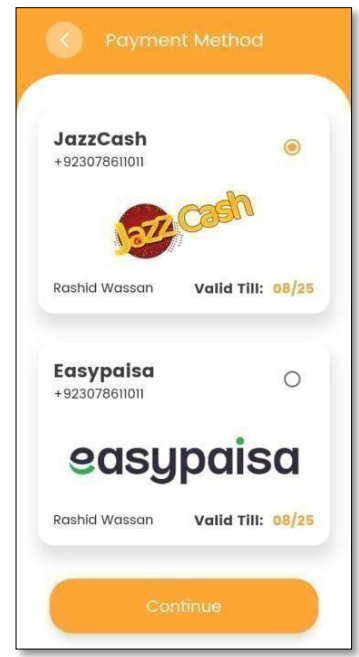

Fig. 9. Payment Method

## V. CONCLUSION

During the phase of the pandemic, people were made to stay alone and follow the SOPs which negatively affected their mentality. It widened the gap in the social life of people that cannot be underestimated and needed to be narrowed down. Also, the need for experts has increased more than ever before and GTMA is proposed to fill this vacuum. The suggested approach intends to provide a system that will let the user find the people of interest and create a consensual meeting both virtually and physically. They can choose the place of their desire and can also fix the meeting in their desired hours. This project not only provides the feasibility to convert boredom into entertainment by meeting new people of your interest but also provides a startup for the human library where people find experts in every field of their area. It will let people meet i.e psychologists, practitioners, doctors, and experienced brains, and have benefited economically. This project also helps society grow better and reduces unemployment as well. Moreover, the education that has been compromised due to the pandemic can also be efficiently restored as it provides feasibility to the students to contact and meet the teachers in their area. Users navigating through the app can see the notification from the subject experts, whom they follow, about their free hours, sessions, and about other location-based events. This saves their time and helps them sort out which session to attend when to fix meetings and whom to meet with. People will meet new people, share their thoughts, and make new friends and contacts. The meetings can be official and formal depending on the book and reader.

The project tends to let people come closer and share their feelings, get counseling, increase their contact, increase their friend circle, and expand their work. It includes many opportunities i. e earning methods for books, knowledge sharing, and counseling, and also provides event bookings. Multiple sessions can be organized and attended. It also reduces unemployment by a considerable level. Moreover, multiple methods of earning will be added not only for books but also for other types of users so that every individual can earn money using this platform and make their time invested.

ACKNOWLEDGEMENT

This research work has been carried out in the Advanced Software Engineering and Research Lab (Department of Computer Systems Engineering, MUET, Jamshoro). The support grant for this research work is covered by National Technology Fund (IGNITE) Federal Government of Pakistan.